\documentclass[10pt]{revtex4}
\usepackage{hyperref}

\begin{document}

\title{ Green's Functions in Perturbative Quantum Gravity}

 \author{ Sudhaker Upadhyay\footnote{e-mail address: sudhakerupadhyay@gmail.com}}
 \affiliation { Department of Physics, Indian Institute of Technology Kanpur, Kanpur 208016, India  }
  \author{
  Bhabani Prasad Mandal\footnote{e-mail address: bhabani.mandal@gmail.com}}

\affiliation { Department of Physics, 
Banaras Hindu University, 
Varanasi 221005, India  }

\begin{abstract}
We show that the Green's functions in non-linear gauge in the theory of perturbative quantum gravity is expressed as a
series in terms of those in linear gauges. This formulation is also holds for operator Green's functions. We further
derive the explicit relation between the Green's functions in the theory of perturbative quantum gravity in  a pair of
arbitary gauges. This process involves some sort of 
modified FFBRST transformations which is derivable from
infinitesimal field-dependent BRST transformations.

 \end{abstract}
  \maketitle 
 
\section{Introduction}

Since its inception,  the general relativity has many striking
similarities to gauge theories. For instance, both involve the idea of local symmetry
and therefore share a number of formal properties. 
 Moreover, consistent quantum gauge theories are well-established
but as yet no satisfactory quantum field theory of gravity has been investigated.
The structures of the Lagrangians of these theories are   rather different. The 
Yang-Mills Lagrangian contains only up to four-point interactions while the
Einstein-Hilbert Lagrangian contains infinitely many interactions.
Despite these differences, string theory provides us sufficient reasons, on the basis of which it can be claimed  that   gravity and gauge theories can, in fact, be unified. 
For example, the Maldacena conjecture \cite{mal,mal1}
relates the weak coupling limit of a gravity theory  
to a strong coupling limit of a special supersymmetric gauge field theory.
 With this similarity, the gauge theories are allowed to be used directly as a resource
  for computations
in perturbative quantum gravity.

 The  perturbative quantum gravity as a gauge theory is a subject of extensive
 research interests \cite{hata1, hata2, asch}. For examples, the mode analysis and Ward identities for a ghost propagator for perturbative quantum gravity  
  has been demonstrated \cite {tsa}.
The Feynman rules and propagator for gravity in the physically interesting cases of inflation have been analysed \cite{wood}.   The propagator for a gauge theory exists only after fixing a gauge.
For instance, the Landau   and   Curci--Ferrari type gauges  have their common uses in the perturbation theory
\cite{dud1, dud2}. Being gauge-fixed, the theory loses their local gauge invariance.
However, it possesses the rather different  the fermionic rigid
 BRST  invariance \cite{brst, tyu}.

The BRST symmetry and the associated concept of BRST cohomology provide the most used covariant quantization method for constrained  systems such as gauge and string theories  \cite{ht,wei}.  
The BRST and the anti-BRST symmetries for perturbative quantum gravity in flat spacetime
  have also been investigated 
\cite{na,ku,ni} which was summarized by N. Nakanishi and I. Ojima \cite{nn}.
  Recently, the BRST formulation for the perturbative quantum gravity
  in general curved spacetime has also  been analyzed  \cite{faiza, upa}. 
  The usual infinitesimal BRST transformation has been generalized by 
  allowing the parameter  finite and field-dependent \cite{sdj}. 
  This FFBRST enjoys the properties of
  usual BRST except it does not leave the path integral measure 
  invariant. The FFBRST transformations have found several applications in gauge field theories 
 in flat spacetime \cite{sdj,sdj1,rb,susk,sb,bss,smm,fs,sud1,sudhak,rbs,das,rs}
 as well as in curved spacetime \cite{sudh}.
The FFBRST formulation to
connect the Green's function of Yang-Mills theory in a set of two otherwise unrelated gauge
choices has been established \cite{sat}.
Nevertheless, the FFBRST formulation to connect Green's functions has not  been developed
so-far in the context of perturbative quantum gravity. The development of FFBRST formulation to connect Green's functions in perturbative quantum gravity is goal of present investigation.

In this paper, we discuss the usual FFBRST transformation in perturbative quantum gravity
to connect the linear and non-linear gauges of the theory.
Further, we establish a connection between arbitrary Green's functions
(or operator Green's functions) in two sets of gauges for the theory of perturbative
quantum gravity. In view of
their extreme importance, we choose these to be the  linear (Landau) and non-linear (Curci-Ferrari) type
gauges.  Here we find that to connect the Green's functions of the theory rather than connection of gauges  
we require different FFBRST transformation.  Finally,  we establish a compact
result expressing an arbitrary Green's function or operator Green's function
in non-linear gauges with a closed expression involving similar Green's functions
in Landau gauges.

This paper is presented as follows. In Sec. II, we  present the  usual FFBRST transformation for a general gauge theory. In Sec. III, we recapitulate the
FFBRST transformation to connect the linear and non-linear gauges in linearized gravity. In Sec. IV, we demonstrate
the similar FFBRST transformation to connect the Green's functions of the
perturbative quantum gravity by a compact formula. 
In the last section, we summarize the results with future motivations.
 \section{The usual FFBRST transformations}
In this subsection, we recapitulate the FFBRST transformation for the general gauge theory in general 
curved spacetime \cite{su}. For this purpose, we first write the usual BRST transformation 
\begin{equation}
\delta_b \phi (x) =s\phi (x) \delta \Lambda,
\end{equation} 
where  $\delta\Lambda$ is  infinitesimal and field-independent
Grassmann parameter and   $\phi (x)$ is the generic notation of fields $(h, c, \bar c, b)$ involved the theory of quantum gravity. The observations of BRST transformation that its
basic properties do not depend on whether 
the parameter $\delta\Lambda$  is (i) finite or infinitesimal, (ii) field-dependent or not, as long 
as it is anticommuting and spacetime independent. This renders
  us a freedom to make the parameter, $\delta\Lambda$ finite and field-dependent without
 affecting its basic features.  The first step towards the goal is to make 
 the  infinitesimal parameter field-dependent by interpolating   a continuous parameter, $\kappa\ (0\leq \kappa\leq 1)$, in the theory.
The fields, $\phi(x,\kappa)$,  depend on  $\kappa$  such that $\phi(x,\kappa =0)=\phi(x)$ is the initial fields and $\phi(x,\kappa 
=1)=\phi^\prime(x)$ is the transformed fields.

The infinitesimal
field-dependent BRST transformation  is defined by \cite{sdj}
\begin{equation}
{d\phi(x,\kappa)}=s  [\phi (x) ]\Theta^\prime [\phi ( \kappa ) ]{d\kappa},
\label{diff}
\end{equation}
where the $\Theta^\prime [\phi ( \kappa ) ]{d\kappa}$ is the infinitesimal but field-dependent parameter.
The FFBRST transformation is then prevailed 
 by integrating this infinitesimal transformation from $\kappa =0$ to $\kappa= 1$,  as follows
\begin{equation}
\phi^\prime\equiv \phi (x,\kappa =1)=\phi(x,\kappa=0)+s [\phi(x) ]\Theta[\phi  ],
\label{kdep}
\end{equation}
where 
\begin{equation}
\Theta [\phi] = \Theta ^\prime [\phi] \frac{ \exp f[\phi]
-1}{f[\phi]},
\label{80}
\end{equation}
 is the finite field-dependent parameter and $f[\phi]$ is given 
 by 
 \begin{eqnarray}
 f[\phi]= \sum_i \int d^4x \frac{ \delta \Theta ^\prime [\phi]}{\delta
\phi_i(x)} s_b \phi_i(x).\label{f}
 \end{eqnarray}  
The resulting FFBRST   transformation  leaves the effective action invariant but the
 functional integral changes non-trivially under it \cite{sdj}. 
 Now we compute  the Jacobian of path integral measure under the FFBRST transformation.

We first  define the Jacobian of the path integral measure under such transformations  with an 
arbitrary finite field-dependent parameter, $\Theta[\phi(x)]$, as
\begin{eqnarray}
{\cal D}\phi' &=&J(
\kappa) {\cal D}\phi(\kappa).
\end{eqnarray}
The Jacobian, $J(\kappa )$,  can be replaced within the functional integral  as
\begin{equation}
J(\kappa )\rightarrow \exp[iS_1[\phi(x,\kappa), \kappa ]],\label{s}
\end{equation}
where $ S_1[\phi (x), \kappa]$ is local functional of fields, iff the following condition gets satisfied  \cite{sdj}
 \begin{eqnarray}
\int {\cal D}\phi   \left[  \frac{1}{J}\frac{dJ}{d\kappa}-i\frac
{dS_1[\phi (x,\kappa ), \kappa]}{d\kappa} \right] e^{i(S_L[\phi ]+S_1[\phi,\kappa ])} =0. \label{mcond}
\end{eqnarray}
 The infinitesimal change in the Jacobian $J(\kappa)$
is addressed with the following formula \cite{sdj}
\begin{equation}
\frac{1}{J}\frac{dJ}{d\kappa}=-\int d^4y\left [\pm s\phi (y,\kappa )\frac{
\delta\Theta^\prime [\phi ]}{\delta\phi (y,\kappa )}\right],\label{jac}
\end{equation}
where sign $+$ is used for bosonic fields $\phi$ and   $-$ sign is used for  fermionic
fields $\phi$.

Recently, exactly similar FFBRST transformations have also been 
considered and general Jacobian is calculated explicitly in terms of
the general finite parameter $\Theta$ \cite{lav}.
\section{The FFBRST transformation in perturbative quantum gravity: Preliminaries } 
In this section we consider perturbative quantum gravity in the framework of FFBRST transformation. In particular 
we analyse the perturbative quantum gravity in linear and non-linear gauges.
Then we generalize the BRST transformation by making the transformation 
finite and field-dependent. Furthermore, we establish the connection between 
these two gauges using FFBRST transformation \cite{su}.
 
   \subsection{The linearized quantum gravity}
 Let us start by writing the classical Lagrangian density for  gravity in general curved spacetime  
\begin{equation} 
{\cal L}_c   = \sqrt{ -g}   (R-2\lambda), \label{kin}
\end{equation}
where $R$ is Ricci scalar curvature and   $\lambda$ is  a cosmological constant.
Here units are setted in such a manner that $16\pi G=1$.
 In the weak approximation the full metric $g_{ab}^f$ can be written as a sum of  fixed
  metric of background spacetime  $g_{ab}$ and the small perturbations around it, denoted by  $h_{ab}$. 
  This fluctuation is considered as a quantum field that needs to be quantized. 
Therefore, numerically
\begin{equation}
g_{ab}^f=g_{ab}+h_{ab}.
\end{equation}
 Incorporating such decomposition, the Lagrangian density given in  (\ref{kin}) described in 
 terms of $h_{ab}$ remains invariant
 under the following  coordinate transformation:
 \begin{eqnarray}
 \delta_\Lambda h_{ab}=\nabla_a \Lambda_b +\nabla_b \Lambda_a + {\pounds}_{(\Lambda)} h_{ab},
 \end{eqnarray}
where the Lie derivative of  $h_{ab}$ with respect to the vector field $\Lambda_a$ is defined  by
 \begin{eqnarray}
 {\pounds}_{(\Lambda)} h_{ab}=\Lambda^c\nabla_c h_{ab} +h_{ac}\nabla_b \Lambda^c
 +h_{ cb}\nabla_a \Lambda^c.
 \end{eqnarray}
The gauge invariance reflects the redundancy in physical  degrees 
 of freedom.  Such redundancy in gauge degrees of 
freedom produces   constraints in the canonical quantization  and leads 
divergences in the generating functional.
 In 
order to fix the 
redundancy we choose the following  gauge-fixing condition  satisfied by quantum field:
 \begin{equation}
G[h]_a=(\nabla^b h_{ab} -\beta\nabla_a h) =0,
\end{equation}
where the parameter $\beta\neq  1$. Because  $\beta=1$ leads to vanishing  conjugate momentum corresponding to $h_{00}$ 
and therefore  generating functional diverges.  
This  gauge-fixing  condition quantum level by adding following term in the classical action: 
  \begin{eqnarray}
{\cal L}_{gf}&=&  \sqrt{- g}[ib^a(\nabla^b h_{ab}-\beta \nabla_a h)]. \label{gfix}
\end{eqnarray} 
The induced (Faddeev--Popov) ghost term is then defined by
\begin{eqnarray}
{\cal L}_{gh}=\sqrt{ -g}\bar c^a M_{ab} c^b,
\end{eqnarray} 
where Faddeev--Popov matrix operator $M_{ab}$ has following expression: 
\begin{equation} 
M_{ab}= i  \nabla_c [ \delta_b^c\nabla_a  + g_{ab}\nabla^c - 2\beta
 \delta_a^c\nabla_b 
+\nabla_b h^c _a -h_{ab}\nabla^c
-h^c_b\nabla_a 
-\beta g^c_ag^{ef}(\nabla_b h_{ef} +h_{eb}\nabla_f +h_{fb}\nabla_e)].
 \end{equation}
Henceforth, the   effective action for perturbative quantum gravity in  curved
spacetime dimensions (in linear gauge) reads
\begin{equation}
S_L =\int d^4 x ({\cal L}_c +{\cal L}_{gf}+{\cal L}_{gh}),  \label{com}
\end{equation}
which  is invariant under following BRST transformations:   
\begin{eqnarray}
s  h_{ab} =   (\nabla_a c_b +\nabla_b c_a +{\pounds}_{(c)} h_{ab}), \  
s c^a  =  -c_b\nabla^b c^a,     \ s  \bar c^a
= b^a,\ s  b^a  =  0.\label{sym}
\end{eqnarray}
Here  we observe that   the gauge-fixing and the ghost parts of the 
effective Lagrangian density  are BRST-exact. Therefore,
\begin{eqnarray}
{\cal L}_g &=& {\cal L}_{gf} +{\cal L}_{gh},\nonumber\\
&=&i s \sqrt{ -g}[\bar c ^a (\nabla^b h_{ab} -\beta\nabla_a h)],\nonumber\\
&=&s\Psi.\label{g}
\end{eqnarray}
The gauge-fixed fermion ($\Psi$) then has the expression
\begin{equation}
\Psi =i  \sqrt{- g}[\bar c ^a (\nabla^b h_{ab} -\beta\nabla_a h)].\label{gff}
\end{equation}
 However,  the 
gauge-fixing and ghost terms in non-linear Curci--Ferrari gauge condition are written by 
\begin{eqnarray}
\mathcal{L'}_g&=& {\cal L'}_{gf} + {\cal L'}_{gh},\nonumber\\
&=&  \sqrt{ -g}\left[ib^a(\nabla^b h_{ab}-\beta \nabla_a h)-i\bar c^b \nabla_b c^a (\nabla^b h_{ab}-
\beta \nabla_a h)+\bar c^a M_{ab} c^b +\frac{\alpha}{2}b^b\nabla_b \bar c^a c_a\right.\nonumber\\
&-&\left. \frac{\alpha}{2} \bar c^c\nabla_c c^b \nabla_b \bar c^a c_a -\frac{\alpha}{2} \bar b^b \nabla_b b^a c_a -\frac{\alpha}{2}\bar c^b\nabla_b \bar c^a c_d\nabla^d c_a -\frac{\alpha}{2} b_a b^a +\alpha \bar c^b b^b \nabla_b c_a\right.\nonumber\\
&+&\left. \alpha \bar c^a \bar c^b c^d \nabla_b \nabla_d c_a\right],\label{nlg}
\end{eqnarray}
where $\alpha$ is a gauge parameter. 
For instance, the effective action,  
having such gauge-fixing and Faddeev--Popov ghost terms,  in non-linear gauge is given by
\begin{equation}
S_{NL} =\int d^4 x ({\cal L}_c +{\cal L'}_{g}), 
\end{equation}
which remains unchanged under following BRST transformations:
\begin{eqnarray}
s \,h_{ab} &=& \nabla_a c_b + \nabla_b c_a + \pounds_{(c)} h_{ab}, \nonumber \\
s \,c^a &=& - c_b \nabla^b c^a, \nonumber \\
s \,\bar{c}^a &=& b^a - \bar{c}^b\nabla_b c^a, \nonumber \\ 
s \,b^a &=& - b^b\nabla_b c^a -  \bar{c}^b c^d\nabla_b \nabla_d c^a.\label{nlbrs}
\end{eqnarray}

   \subsection{FFBRST transformation for linear to non-linear gauge}
  We construct the FFBRST transformation   for perturbative quantum gravity 
  utilizing   the BRST transformation (\ref{sym}) as follows
\begin{eqnarray}
f \,h_{ab} &=& (\nabla_a c_b + \nabla_b c_a + \pounds_{(c)} h_{ab})\ \Theta[\phi], \nonumber \\
f \,c^a &=& - c_b \nabla^b c^a\ \Theta[\phi], \nonumber \\
f \,\bar{c}^a &=&  b^a   \ \Theta[\phi], \nonumber \\ 
f \,b^a &=& 0,\label{ff}
\end{eqnarray}
where $ \Theta[\phi]$ is an arbitrary finite field-dependent parameter. To establish the connection 
between  the Landau and the (non-linear) Curci--Ferrari gauge we opt the finite field-dependent parameter
constructed from following infinitesimal field-dependent parameter:
\begin{equation}
\Theta'[\phi] = i\frac{\alpha}{2}\sqrt{-g}\int d^4 y\ (\bar c_b \nabla^b\bar c ^a c_a -\bar c^a b_a -\bar c^a 
\bar c_b \nabla^b c_a).\label{theta}
\end{equation}
Exploiting relations (\ref{jac}) and (\ref{theta})  we calculate the change in Jacobian  as
\begin{eqnarray}
\frac{1}{J(\kappa)}\frac{dJ(\kappa)}{d\kappa}&=& -i\frac{\alpha}{2}\sqrt{-g}\int d^4x \left[
-b_b\nabla^b \bar c^a c_a + \bar c^d \nabla_d c_b\nabla^b\bar c^a c_a
+\bar c_b \nabla^b b^a c_a +\bar c_b \nabla^b \bar c^a c_d\nabla^d c_a\right.
\nonumber\\
&+&\left. b_a b^a -2\bar c^a b_b\nabla^b c_a
-2\bar c^a \bar c_b c_d \nabla^b\nabla^d c_a\right].\label{j}
\end{eqnarray}
The local functional $S_1$  in the expression (\ref{s})
is written by
\begin{eqnarray}
S_1[\phi(\kappa), \kappa]&=& \int d^4x \left[\xi_1 
 b_b\nabla^b \bar c^a c_a +\xi_2 \bar c^d \nabla_d c_b\nabla^b\bar c^a c_a
+\xi_3\bar c_b \nabla^b b^a c_a +\xi_4\bar c_b \nabla^b \bar c^a c_d\nabla^d c_a\right.
\nonumber\\
&+&\left.\xi_5 b_a b^a +\xi_6 \bar c^a b_b\nabla^b c_a
+\xi_7 \bar c^a \bar c_b c_d \nabla^b\nabla^d c_a\right],
\end{eqnarray}
where parameters $\xi_i (i=1,2,..,7)$  depend explicitly on parameter $\kappa$ as follows  \cite{su},
\begin{eqnarray}
&&\xi_1  = 
\frac{\alpha}{2} \sqrt{-g}\kappa,\ \ \xi_2 = -\frac{\alpha}{2} \sqrt{-g} \kappa,\ \ \xi_3 = -\frac{\alpha}{2} \sqrt{-g}\kappa,\ \ \xi_4
= -\frac{\alpha}{2} \sqrt{-g} \kappa,\nonumber\\
&& \xi_5 =- \frac{\alpha}{2} \sqrt{-g} \kappa,\ \ \xi_6  = \alpha \sqrt{-g} \kappa,\ \ \xi_7 =  \alpha \sqrt{-g} \kappa. 
\end{eqnarray}
With these identifications of $\xi_i (\kappa)$ 
the expression of $S_1$ becomes
\begin{eqnarray}
S_1[\phi(\kappa), \kappa]&=&   \kappa\int d^4x  \sqrt{-g}\left[  \frac{\alpha}{2}  
 b_b\nabla^b \bar c^a c_a -\frac{\alpha}{2} \bar c^d \nabla_d c_b\nabla^b\bar c^a c_a
-\frac{\alpha}{2}\bar c_b \nabla^b b^a c_a -\frac{\alpha}{2} \bar c_b \nabla^b \bar c^a c_d\nabla^d c_a\right.
\nonumber\\
&-&\left.\frac{\alpha}{2} b_a b^a +\alpha \bar c^a b_b\nabla^b c_a
+\alpha \bar c^a \bar c_b c_d \nabla^b\nabla^d c_a\right].
\end{eqnarray}
Therefore, the FFBRST transformation (\ref{ff}) 
changes 
the
effective action within functional integration as
\begin{eqnarray}
S_{L} +S_1 (\kappa =1) &=&\int d^4 x \left[{\cal L}_c +i \sqrt{-g}b^a(\nabla^b h_{ab}-\beta \nabla_a h)+\sqrt{-g}\bar c^a M_{ab} c^b \right.\nonumber\\
&+&\left. \frac{\alpha}{2}  \sqrt{-g}
 b_b\nabla^b \bar c^a c_a -\frac{\alpha}{2}\sqrt{-g} \bar c^d \nabla_d c_b\nabla^b\bar c^a c_a
-\frac{\alpha}{2}\sqrt{-g}\bar c_b \nabla^b b^a c_a -\frac{\alpha}{2} \sqrt{-g}\bar c_b \nabla^b \bar c^a c_d\nabla^d c_a\right.
\nonumber\\
&-&\left.\frac{\alpha}{2}\sqrt{-g} b_a b^a +\alpha \sqrt{-g}\bar c^a b_b\nabla^b c_a
+\alpha \sqrt{-g}\bar c^a \bar c_b c_d \nabla^b\nabla^d c_a\right].
\end{eqnarray}
After performing a shift in the Nakanishi--Lautrup field by $\bar c^b \nabla_b c^a $, the above expression reduces to
\begin{eqnarray}
S_{L} +S_1 (\kappa =1) &=&\int d^4 x \left[{\cal L}_c +i \sqrt{-g}b^a(\nabla^b h_{ab}-\beta
 \nabla_a h)-i \sqrt{-g}\bar c^b \nabla_b c^a (\nabla^b h_{ab}-\beta \nabla_a h)  \right.\nonumber\\
&+&\left.\sqrt{-g}\bar c^a M_{ab} c^b \right],
\nonumber\\
&=&S_{NL},
\end{eqnarray}
which is nothing but the effective action for perturbative quantum gravity in Landau gauge. 
\section{Relation between Green's function for linear and non-linear gauges }
In this section, we establish a   procedure for   FFBRST transformation  that
transforms the generating functional (Green's function) in one kind of a gauge choice to the generating functional in another kind of a
gauge choice. For this purpose we define
the generating functional for perturbative quantum gravity in linear gauge
\begin{eqnarray}
W_L=\int {\cal D}\phi\ e^{iS_L[\phi]},
\end{eqnarray}
which transforms under FFBRST transformation $\phi'(x)=\phi(x)+s\phi\Theta[\phi]$ defined in (\ref{ff})
as follows:
\begin{eqnarray}
W_{NL}=\int {\cal D}\phi'\ e^{iS_L[\phi']}=W_L.
\end{eqnarray}
Now, we  want to implement this transformation
to connect  the Green’s functions in the two gauges for quantum gravity theory.  
According to the standard procedure, $n$-point Green's functions in non-linear gauge under FFBRST transformation
transform as 
\begin{eqnarray}
G^{NL}_{i_1....i_n}&=&\int {\cal D}\phi'\prod_{r=1}^n\phi'_{i_r}e^{iS_{NL}[\phi']},\nonumber\\
&=& \int {\cal D}\phi \prod_{r=1}^n(\phi_{i_r}+s_{i_r}\phi\Theta[\phi])e^{iS_{L}[\phi']},\nonumber\\
&=& G^{L}_{i_1....i_n}+\Delta G^{L}_{i_1....i_n},
\end{eqnarray}
where $\Delta G^{L}_{i_1....i_n}$, refers
the difference between the $n$-point Green's functions  in the two sets
of gauges. This   may involve additional vertices
corresponding to insertions of operators $s_{i_r}\phi$. But it seems technically incorrect for the
following reasons.

A priory, it is not obvious that if condition (\ref{mcond}) (for replacing Jacobian to $e^{iS_1}$)  holds
for quantum gravity then an equation  modified  to include an arbitrary  operator ${\cal O}[\phi]$
of type
 \begin{eqnarray}
\int {\cal D}\phi  {\cal O}[\phi] \left[  \frac{1}{J}\frac{dJ}{d\kappa}-i\frac
{dS_1[\phi (x,\kappa ), \kappa]}{d\kappa} \right] e^{i(S_L[\phi ]+S_1[\phi,\kappa ])} =0,\label{cond}
\end{eqnarray}
  would also hold.
  Of course it does not hold in general for the reason discussed in \cite{sat}.
  For this reason,
to connect the Green's functions for the two type of
gauges we need a elaborate  treatment of FFBRST transformation.

We begin with a general Green's function in non-linear gauge defined by
\begin{eqnarray}
G=\int {\cal D}\phi'{\cal O}[\phi'] e^{iS_{NL}[\phi']},\label{gre}
\end{eqnarray}
where ${\cal O}[\phi']$ is an arbitrary operator. So, (\ref{gre}) covers both the arbitrary operator
Green's functions as well as arbitrary ordinary Green's functions.
Specifically, for ${\cal O}_1[\phi']=h'_{ab}h'_{cd}$ (\ref{gre}) describes the gauge graviton propagator,
however, for ${\cal O}_2[\phi']=h'_{ab}\bar c'^{c} c'_c$ it describes the 3-point propagator.
We want to express the Green's function ($G$) of perturbative gravity entirely in terms
of the linear type gauge Green's functions (and possibly involving vertices
from $s\phi$). So we define
  \begin{eqnarray}
G(\kappa)=\int {\cal D}\phi  {\cal O}[\phi(\kappa),\kappa]   e^{i(S_L[\phi ]+S_1[\phi,\kappa ])},\label{mod}
\end{eqnarray}
 where the form of operator $ {\cal O}[\phi(\kappa),\kappa]$ demands
 \begin{eqnarray}
 \frac{dG}{d\kappa}=0.\label{G}
 \end{eqnarray}
Under FFBRST transformation ($\kappa=1$), it reflects that
\begin{eqnarray}
G(1)=\int {\cal D}\phi'  {\cal O}[\phi',1]   e^{i S_{NL}[\phi' ] },\label{g1}
\end{eqnarray}
which coincides with (\ref{gre}), where as
at $\kappa=0$ this reads
\begin{eqnarray}
G(0)=\int {\cal D}\phi   {\cal O}[\phi,0]   e^{i S_{L}[\phi ] },
\end{eqnarray}
and is numerically equal to (\ref{g1}).
Now, we  need to determine the form of  ${\cal O}[\phi(\kappa),\kappa]$ in
(\ref{mod})   so that the condition (\ref{G}) gets satisfied. For this purpose, 
we perform the field transformation from $\phi(\kappa)$ to $\phi(\kappa+d\kappa)$
through infinitesimal field-dependent BRST transformation defined in (\ref{diff}) which leads
\begin{eqnarray}
G(\kappa)&=&\int {\cal D}\phi (\kappa+d\kappa)\frac{J(\kappa+d\kappa)}{J(\kappa)} \left({\cal O}[\phi(\kappa+d\kappa),\kappa+d\kappa]-s\phi\Theta'\frac{\delta{\cal O}}{\delta\phi}d\kappa +
\frac{\partial{\cal O}}{\partial \kappa} d\kappa\right)\times\nonumber\\
&&\left(1-i\frac{dS_1}{d\kappa}d\kappa \right)e^{i S_{L}[\phi(\kappa+d\kappa)] +iS_1[\phi(\kappa+d\kappa),\kappa+d\kappa ]},\nonumber\\
&=&\int {\cal D}\phi (\kappa+d\kappa)\left(1+\frac{1}{J}\frac{dJ}{d\kappa}d\kappa \right) \left({\cal O}[\phi(\kappa+d\kappa),\kappa+d\kappa]-s\phi\Theta'\frac{\delta{\cal O}}{\delta\phi}d\kappa +
\frac{\partial{\cal O}}{\partial \kappa} d\kappa\right)\times\nonumber\\
&&\left(1-i\frac{dS_1}{d\kappa}d\kappa \right)e^{i S_{L}[\phi(\kappa+d\kappa)] +iS_1[\phi(\kappa+d\kappa),\kappa+d\kappa ]},\nonumber\\
&=& G(\kappa+d\kappa),
\end{eqnarray} 
iff
\begin{eqnarray}
\int {\cal D}\phi (\kappa)\left(\left[ \frac{1}{J}\frac{dJ}{d\kappa}-i\frac{dS_1}{d\kappa}\right] {\cal O}
[\phi(\kappa),\kappa]-s\phi\Theta'\frac{\delta{\cal O}}{\delta\phi} +
\frac{\partial{\cal O}}{\partial \kappa} \right)  e^{i S_{L}[\phi(\kappa)] +iS_1[\phi(\kappa),\kappa ]}=0.
\label{mcond0}
\end{eqnarray}
So we get precisely correct expression (\ref{mcond0}) for replacing Jacobian of path integral measure in 
Green's function of quantum gravity
as $e^{iS_1}$ in place of incorrect one (\ref{cond}).

Exploiting the information of above expression, the required condition for $\kappa$-independence of $G$
is
\begin{eqnarray}
&&\int {\cal D}\phi(\kappa)
e^{i S_{L}[\phi(\kappa)] +iS_1[\phi(\kappa),\kappa ]}\left( \frac{\partial {\cal O}}{\partial \kappa}
+\int (\nabla_a c_b +\nabla_b c_a +{\pounds}_{(c)} h_{ab})\Theta' \frac{\delta {\cal O}}{\delta h_{ab}}
-\int c_b\nabla^b c_a \Theta' \frac{\delta {\cal O}}{\delta c_{a}}\right.\nonumber\\
&&\left.+\int \left[b_a - \kappa\bar{c}^b\nabla_b c^a\right]  \Theta'\frac{\delta {\cal O}}{\delta \bar c_a}\right) =0.\label{sss}
\end{eqnarray}
Now, if we construct
the operator ${\cal O}$ to satisfy 
\begin{eqnarray}
&&   \frac{\partial {\cal O}}{\partial \kappa}
+\int (\nabla_a c_b +\nabla_b c_a +{\pounds}_{(c)} h_{ab})\Theta' \frac{\delta {\cal O}}{\delta h_{ab}}
-\int c_b\nabla^b c_a \Theta' \frac{\delta {\cal O}}{\delta c_{a}} \nonumber\\
&& +\int \left[b_a - \kappa\bar{c}^b\nabla_b c^a\right]  \Theta'\frac{\delta {\cal O}}{\delta \bar c_a}  =0.
\label{o}
\end{eqnarray}
Then condition (\ref{sss}) automatically gets  satisfied.
Now, we consider a new set of fields ($\tilde{h}_{ab}, \tilde{c}_a, \tilde{\bar{c}}_a, \tilde{b}_a$)
having following infinitesimal field-dependent BRST transformation:
\begin{eqnarray}
\frac{\delta \tilde h_{ab} }{\delta \kappa}  &=& (\nabla_a \tilde c_b + \nabla_b\tilde c_a + \pounds_{(\tilde c)}\tilde h_{ab})\ \Theta'[\tilde\phi], \nonumber \\
\frac{\delta \tilde c^a} {\delta \kappa}  &=& -\tilde c_b \nabla^b\tilde c^a\ \Theta'[\tilde\phi], \nonumber \\
\frac{\delta \tilde{\bar{c}}^a}{\delta \kappa} &=& \tilde B^a   \ \Theta'[\tilde\phi], \nonumber \\ 
\frac{\delta \tilde B^a }{\delta \kappa} &=& 0, \label{ifi}
\end{eqnarray}
where $\tilde B^a= \tilde b^a-\kappa \tilde{\bar{c}}^b\nabla_b \tilde c^a$.
These new fields satisfy the following boundary condition: $\tilde{\phi(1)}=\phi(1)$.
The condition (\ref{o}) for  ${\cal O}[\tilde\phi (\kappa),\kappa]$ instead of ${\cal O}[\phi (\kappa),\kappa]$ reads
\begin{equation}
\frac{d{\cal O}[\tilde\phi (\kappa),\kappa]}{d\kappa}=0.
\end{equation}
Now utilizing $
{\cal O}[\tilde\phi (1),1]={\cal O}[ \phi (1),1] ={\cal O}[ \phi']
$ we obtain
\begin{eqnarray}
{\cal O}[\tilde\phi (\kappa),\kappa]={\cal O}[ \phi'],\label{ko}
\end{eqnarray}
which  tells us how the operator ${\cal O}[\phi (\kappa),\kappa]$ evolves.
To derive FFBRST transformation corresponding to (\ref{ifi}),
we first define the modification in $f$ of (\ref{f}) as follows,
\begin{eqnarray}
 {f}[\tilde{\phi},\kappa]=f_1[\tilde{\phi}]+\kappa f_2[\tilde{\phi}].
\end{eqnarray}
Therefore,
\begin{eqnarray}
\frac{d\Theta'[\tilde{\phi}(\kappa)]}{d\kappa} =(f_1[\tilde{\phi}]+\kappa f_2[\tilde{\phi}])\Theta'[\tilde{\phi}(\kappa)]
\end{eqnarray}
Performing integration from $0$ to $\kappa$,
\begin{eqnarray}
\Theta'[\tilde{\phi}(\kappa)] =\Theta[\phi]\exp
\left(\kappa f_1[ {\phi}]+\frac{\kappa^2}{2} f_2[{\phi}]\right).
\end{eqnarray}
Similarly, integrating (\ref{ifi}) we get
the FFBRST transformation, written compactly as,
\begin{eqnarray}
\phi' &=&\phi+\left[(\tilde{\delta}_1[\phi] +\tilde{\delta}_2[\phi] )\int d\kappa\exp
\left(\kappa f_1[ {\phi}]+\frac{\kappa^2}{2} f_2[{\phi}]\right)\right]\Theta'[\phi],\nonumber\\
&=&\phi+\delta\phi[\phi].\label{fifi}
\end{eqnarray}
 Now we apply FFBRST transformation (\ref{fifi}) on Green's function in non-linear gauge (\ref{gre})
\begin{eqnarray}
G&=& \int {\cal D}\phi'{\cal O}[\phi'] e^{iS_{NL}[\phi']},\nonumber\\
&=& \int {\cal D}\phi {\cal O}[\phi ++\delta\phi[\phi] ] e^{iS_{ L}[\phi]},\nonumber\\
&=& \int {\cal D}\phi {\cal O}[\phi] e^{iS_{ L}[\phi]}\nonumber\\
&+&\int {\cal D}\phi   \left[(\tilde{\delta}_1[\phi] +\tilde{\delta}_2[\phi] )\int d\kappa\exp
\left(\kappa f_1[ {\phi}]+\frac{\kappa^2}{2} f_2[{\phi}]\right)\right]\Theta' [\phi] \frac{\delta{\cal O}[\phi ]}{\delta\phi}e^{iS_{ L}[\phi]}.
\end{eqnarray}

Further, it can be written by
\begin{eqnarray}
\langle {\cal O}\rangle_{NL}=\langle {\cal O}\rangle_{L}+\int_0^1 d\kappa\int{\cal D}\phi (\tilde\delta_1[\phi] +\tilde\delta_2 [\phi])\Theta'[\phi]\frac{\delta{\cal O}[\phi ]}{\delta\phi}e^{iS_M},
\end{eqnarray}
where $iS_M= =iS_L+\kappa f_1[ {\phi}]+\frac{\kappa^2}{2} f_2[{\phi}]$
In this way, we establish the connection between the Green's function in two gauges 
in perturbative quantum gravity.
\section{Concluding remarks}
In this work, unlike to the usual FFBRST transformation we have demonstrated the
different FFBRST transformation in case of perturbative quantum gravity
to relate the arbitrary Green's functions of the  theory corresponding to two different gauges. 
 For concreteness, we have considered the linear and the non-linear 
gauges from the point of view of their common usage in gravity theory.
The Green's functions in non-linear gauge in the theory of perturbative quantum gravity is expressed as a
series in terms of those in linear gauges.
In this context we have shown the remarkable difference between the
 the modified FFBRST transformation and the usual one. Further, being related to the usual FFBRST formulation,  this  modified FFBRST transformation is
 obtained by integration of (\ref{ifi}).
 We hope
that the final result  putted in a simple form    will be very useful from computational  point of view in the theory of perturbative quantum gravity.


\begin{thebibliography}{99}

\bibitem{mal} J. Maldacena,  
Int. J. Theor. Phys.
38, 1113 (1998).
\bibitem{mal1} O. Aharony, S. S. Gubser, J. Maldacena, H. Ooguri and Y. Oz,  Phys. Rept. 323, 183 (2000).


 
\bibitem{hata1} T. Hatanaka and S. V. Ketov,  Nucl. Phys. B 794, 495 (2008).
\bibitem{hata2}  T. Hatanaka and S. V. Ketov, Class. Quant. Grav. 23, L45 (2006).
\bibitem{asch}  P. Aschieri, M. Dimitrijevic, F. Meyer and J. Wess, Class. Quant. 
Grav. 23, 1883 (2006).
\bibitem{tsa}  N. C. Tsamis and R. P. Woodard, Phys. Lett.  B 292,  269 (1992). 
 
\bibitem{wood} J. Iliopoulos, T. N. Tomaras, N. C. Tsamis and R. P. Woodard, Nucl. Phys. B 534,  419 (1998).
 
\bibitem{dud1}D. Dudal, H. Verschelde, V. E. R. Lemes, M. S. Sarandy, S. P. Sorella and M. Picariello, 
Ann. Phys. 308, 62 (2003).
\bibitem{dud2}D. Dudal, V. E. R. Lemes, M. Picariello, M. S. Sarandy, S. P. Sorella and H. Verschelde, 
JHEP. 0212, 008 (2002).  
\bibitem{brst} C. Becchi, A. Rouet and R. Stora,  Annals Phys. {98}, 287 
(1974). 
\bibitem {tyu} I. V. Tyutin, Lebedev Physics Institute preprint 39 (1975), arXiv: 0812.0580.
\bibitem{ht} M. Henneaux and C. Teitelboim,  Quantization of gauge
systems  (Princeton, USA: Univ. Press, 1992).
\bibitem{wei} S. Weinberg,   The quantum theory of fields, Vol-II: Modern
applications (Cambridge, UK Univ. Press, 1996).


\bibitem{na} N. Nakanishi, Prog. Theor. Phys. 59, 972 (1978).
\bibitem{ku} T. Kugo and I. Ojima, Nucl. Phys. B 144, 234 (1978).
\bibitem {ni} K. Nishijima and M. Okawa, Prog. Theor. Phys. 60, 272 (1978).
\bibitem{nn} N. Nakanishi and I. Ojima, Covariant operator formalism of gauge theories
and quantum gravity  (World Sci. Lect. Notes. Phys.  1990).
 
\bibitem{faiza} M. Faizal, Found. Phys. 41, 270  (2011).
\bibitem{upa} S. Upadhyay,  Phys. Lett. B 723, 470 (2013); Annls.  Phys. 344, 290  (2014). 
  \bibitem{sdj} S. D. Joglekar and B. P. Mandal,   {Phys. Rev.}  {D 51}, 1919 (1995).
\bibitem{sdj1}  S. D. Joglekar and B. P. Mandal, Int. J. Mod. Phys. A 17, 1279 (2002).
\bibitem{rb} R. Banerjee and B. P. Mandal, Phys. Lett. B 488, 27 (2000).
 \bibitem{susk}   S. Upadhyay,   S. K. Rai and B. P. Mandal,  J. Math. Phys.  {52}, {022301} (2011).
\bibitem{sb} S. Upadhyay and B. P. Mandal,  arXiv:1409.1735 [hep-th];
 arXiv:1407.2017 [hep-th];  Prog. Theor. Exp. Phys.  053B04 (2014);  Eur. Phys. J.  {C 72},  2065 
(2012); Annls.   Phys. {327}, 2885 (2012); EPL  {93}, 
{31001} (2011); Mod. Phys. Lett.   {A 25}, {3347} (2010); arXiv:1503.07390.   
\bibitem{bss} B. P. Mandal, S. K. Rai, and S. Upadhyay, EPL 
92, 21001 (2010).  
\bibitem{smm} S. Upadhyay, M. K. Dwivedi and B. P. Mandal, Int. J. Mod. Phys. A 28, 1350033 (2013).
\bibitem{fs} M. Faizal, B. P. Mandal and S. Upadhyay, Phys. Lett. B 721, 159 (2013).
\bibitem{sud1}  R. Banerjee, B. Paul and S. Upadhyay,  Phys. Rev. D 88, 065
019 (2013).
\bibitem{sudhak} S. Upadhyay,  EPL 105, 21001  (2014); Phys. Lett. B 727, 293 (2013);   EPL 104, 61001  (2013);  arXiv:1308.0982 [hep-th].
\bibitem{rbs} R. Banerjee, B. Paul, S. Upadhyay, Phys. Rev. D 88 (2013)  065019.
\bibitem{das} S. Upadhyay and D. Das,  Phys. Lett. B 733, 63 (2014).
\bibitem{rs} R. Banerjee and S. Upadhyay,  Phys. Lett. B 734, 369 (2014).
\bibitem{sudh} S. Upadhyay, Annls.   Phys. 356, 299 (2015);  Phys. Lett. B 740, 341 (2015).
\bibitem{hig} A. Higuchi and S. S. Kouris, Class. Quant. Grav. 18, 4317 (2001).
 
 \bibitem{sat}  S. D. Joglekar and A. Mishra, J. Math. Phys. 41, 1755 (2000).
 \bibitem{su} S. Upadhyay,  Annls.   Phys. 340, 110  (2014).
 
\bibitem{lav} P. M. Lavrov and O. Lechtenfeld, Phys. Lett. B 725, 382 (2013).
\end{thebibliography}
\end{document}